\documentclass[aps,prl,twocolumn,groupedaddress,showpacs,reprint]{revtex4-1}

\usepackage{amsmath}    
\usepackage{amsfonts}
\usepackage{bm}
\usepackage{amssymb}
\usepackage{graphicx}   
\usepackage{subfigure}  
\usepackage{comment}

\begin{document}

\title{Exceptional contours and band structure design in parity-time symmetric photonic crystals}


\author{Alexander Cerjan, Aaswath Raman, and Shanhui Fan}
\affiliation{Department of Electrical Engineering, and Ginzton Laboratory, Stanford  University,  Stanford,  California  94305,  USA}

\date{\today}

\begin{abstract}
We investigate the properties of multidimensional parity-time symmetric periodic systems whose non-Hermitian periodicity
is an integer multiple of the underlying Hermitian system's periodicity. This creates a natural set of degeneracies which can undergo
thresholdless $\mathcal{PT}$ transitions. 
We derive a $\mathbf{k} \cdot \mathbf{p}$ perturbation theory suited to the
continuous eigenvalues of such systems in terms of the modes of the underlying Hermitian system. 
In photonic crystals, such thresholdless $\mathcal{PT}$ transitions are shown to yield significant control over the
band structure of the system, and can result in all-angle supercollimation, a $\mathcal{PT}$-superprism
effect, and unidirectional behavior.
\end{abstract}


\maketitle

Over the past few years, there has been substantial interest in the properties of parity-time
symmetric optical systems, which are invariant under the combined action
of parity ($\mathcal{P}$) and time reversal ($\mathcal{T}$) operations. These systems exhibit a completely real spectrum despite
their non-Hermitian nature until a phase transition occurs at an exceptional point, beyond which at least
two of the system's eigenvalues leave the real axis and become a complex-conjugate pair \cite{bender_pt-symmetric_1999,bender_complex_2002}. 
The motivation for studying such systems originally stemmed from the equivalence
of the Schr\"{o}dinger equation with the paraxial wave equation, and early
work focused on demonstrating the $\mathcal{PT}$ transition in optical
waveguide arrays \cite{musslimani_optical_2008,makris_beam_2008,klaiman_visualization_2008,longhi_bloch_2009,makris_pt-symmetric_2010,ruter_observation_2010,szameit_pt-symmetry_2011}.
Subsequently, many other systems which display intriguing and counter-intuitive phenomena due to the presence of
an exceptional point have been discovered, such as $\mathcal{PT}$ symmetric optical
cavities \cite{chong_pt-symmetry_2011,ge_conservation_2012}, loss-induced
transmission through optical waveguides \cite{guo_observation_2009}, unidirectional reflection or
transmission \cite{lin_unidirectional_2011,regensburger_parity-time_2012,feng_experimental_2013,peng_parity-time-symmetric_2014,chang_parity-time_2014}, 
and lasers with reversed pump dependence \cite{liertzer_pump-induced_2012,brandstetter_reversing_2014,peng_loss-induced_2014} or enhanced
single-mode behavior \cite{hodaei_parity-time_symmetric_2014,feng_single-mode_2014}. 

Recently it has been recognized that there are two types of $\mathcal{PT}$ transitions: 
ordinary $\mathcal{PT}$
transitions wherein the entire spectrum remains real up to a non-zero threshold amount of gain and loss,
and thresholdless $\mathcal{PT}$ transitions for which an infinitesimal amount of gain and loss
yields a complex spectrum \cite{ge_parity-time_2014}.  
While ordinary $\mathcal{PT}$ transitions
exist in all $\mathcal{PT}$ symmetric systems, thresholdless $\mathcal{PT}$ transitions
require that the underlying Hermitian system both possesses a degeneracy, and that
the addition of non-Hermitian material couples the degenerate modes. 
However, with a few exceptions \cite{szameit_pt-symmetry_2011,ge_lieb_lattice_arxiv}, most previous work on $\mathcal{PT}$ symmetric systems has been confined to one-dimensional
or quasi-1D systems, which have not possessed such generic degeneracies.
Furthermore, previous studies of higher-dimensional periodic parity-time symmetric structures have 
imposed the $\mathcal{PT}$ perturbation within every primitive cell of the Hermitian system separately \cite{musslimani_optical_2008,makris_beam_2008,szameit_pt-symmetry_2011}, which does not 
guarantee a thresholdless $\mathcal{PT}$ transition exists.
Very recently, an accidental degeneracy in the band structure of
a two-dimensional photonic crystal slab was used to create
a thresholdless $\mathcal{PT}$ transition  \cite{zhen_spawning_2015}.
However, creating an accidental degeneracy requires careful engineering of the crystal
geometry.

In this Letter, we introduce a general mechanism for realizing thresholdless $\mathcal{PT}$ transitions. 
We consider multi-dimensional $\mathcal{PT}$ symmetric photonic crystals (PhC), whose non-Hermitian primitive cell is 
an integer multiple of the primitive cell of the underlying Hermitian system. We show that under a 
very general set of conditions, such systems always exhibit a thresholdless $\mathcal{PT}$ transition in part of the 
wavevector space. This yields an important experimental
advantage, by enabling access to thresholdless $\mathcal{PT}$ transitions, the strength of gain and loss
required to observe $\mathcal{PT}$ 
phenomena is greatly reduced.
Moreover, such a system enables a new form of band structure engineering, and can result in a
$\mathcal{PT}$-superprism effect, unidirectional behavior, and all-angle supercollimation, which are distinct
from related effects in Hermitian PhCs \cite{joannopoulos,kosaka_superprism_1998,kosaka_superprism_1999,luo_all-angle_2002,yu_bends_2003}.

\begin{figure}[!ht]
    \centering
    \subfigure{
    \centering
        \includegraphics[width=0.23\textwidth]{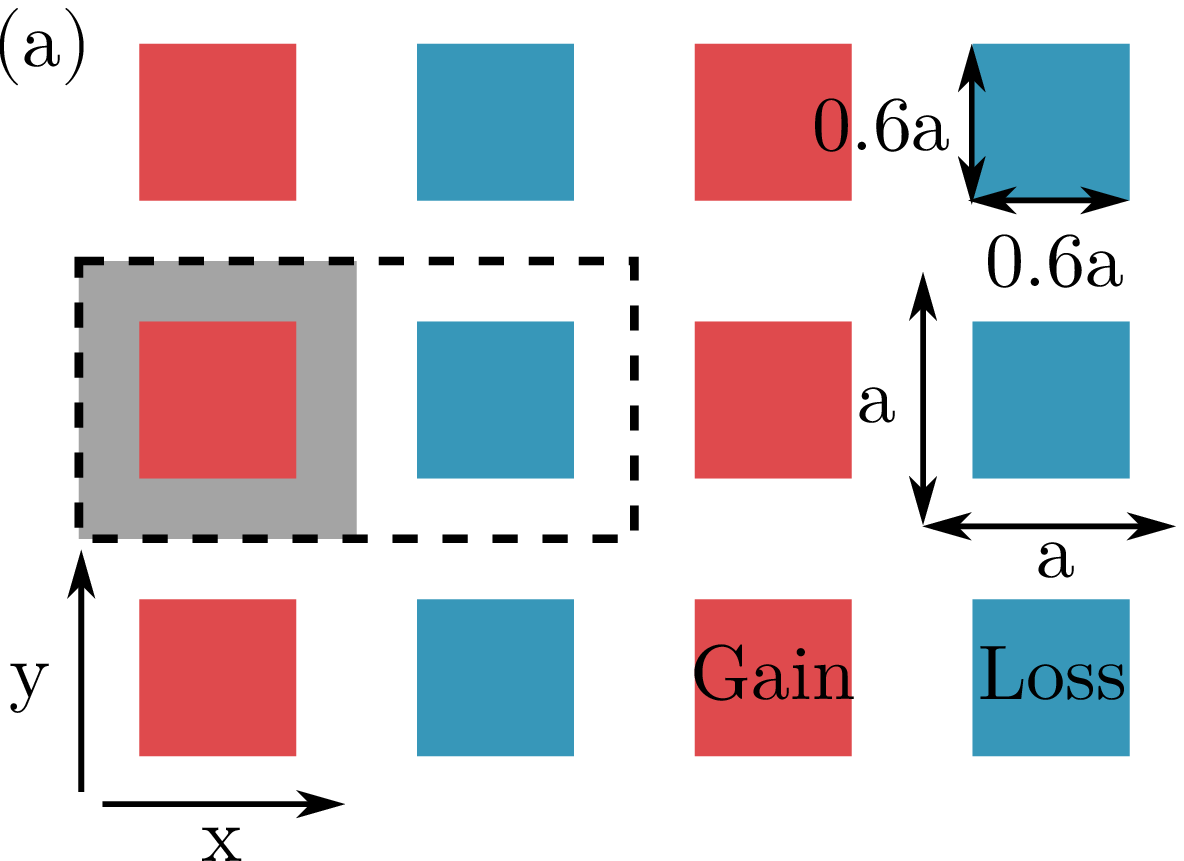}
        \includegraphics[width=0.23\textwidth]{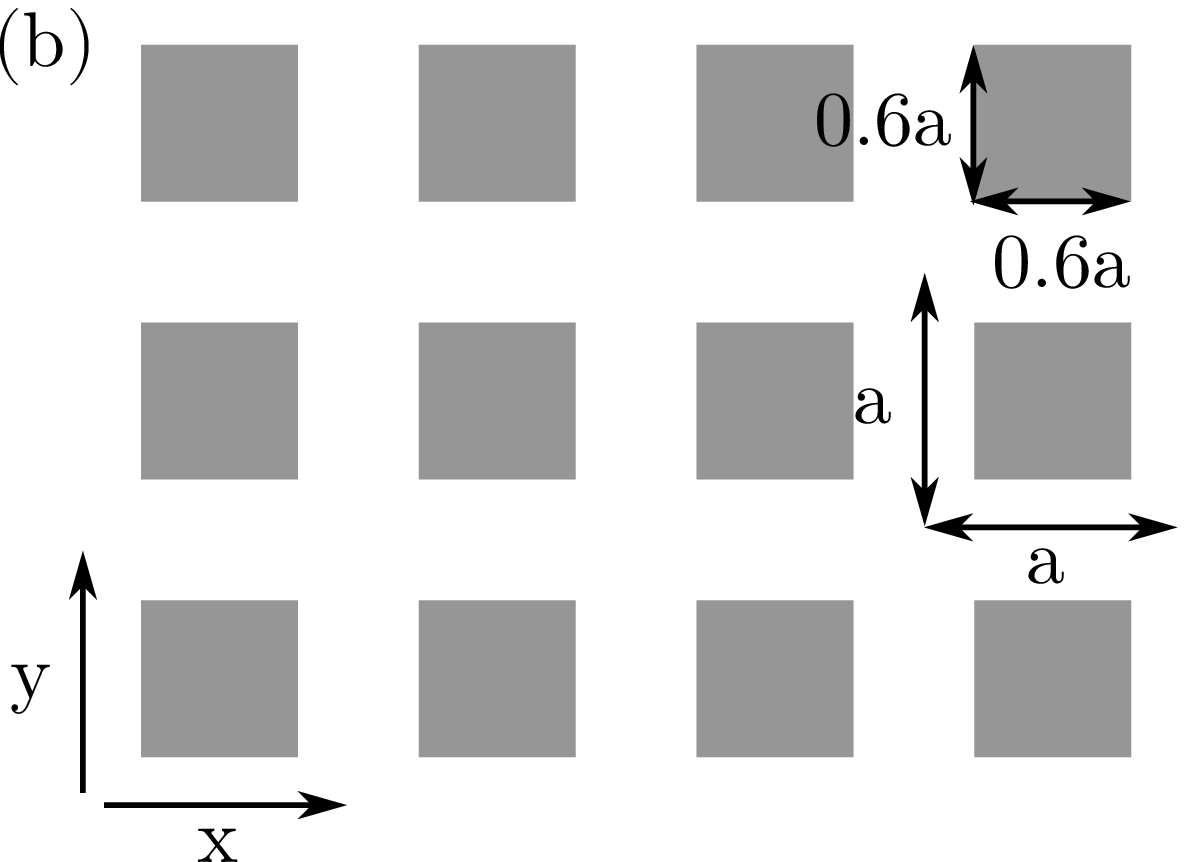}
    }
    \subfigure{
    \centering
        \includegraphics[width=0.22\textwidth]{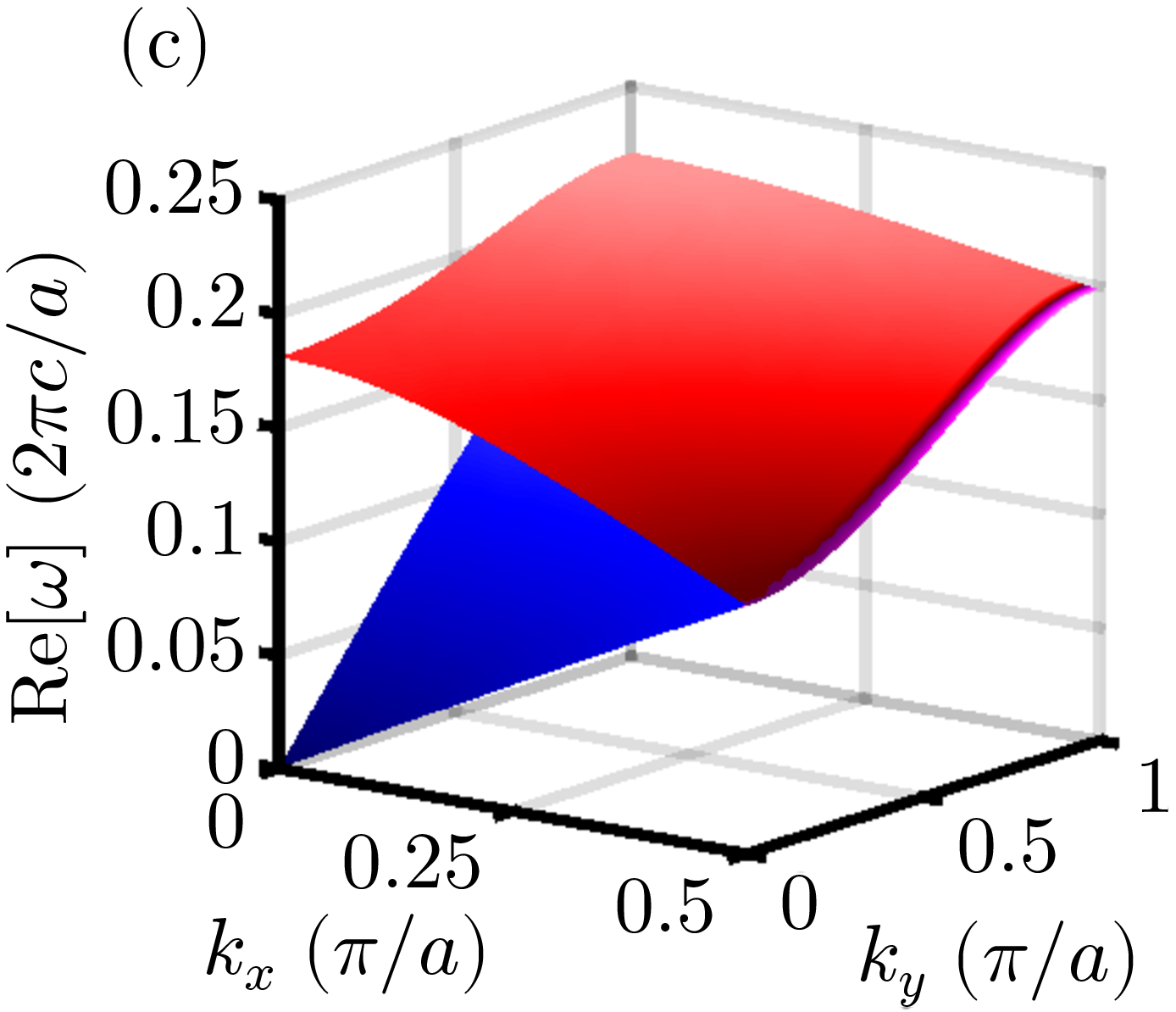}
        \includegraphics[width=0.22\textwidth]{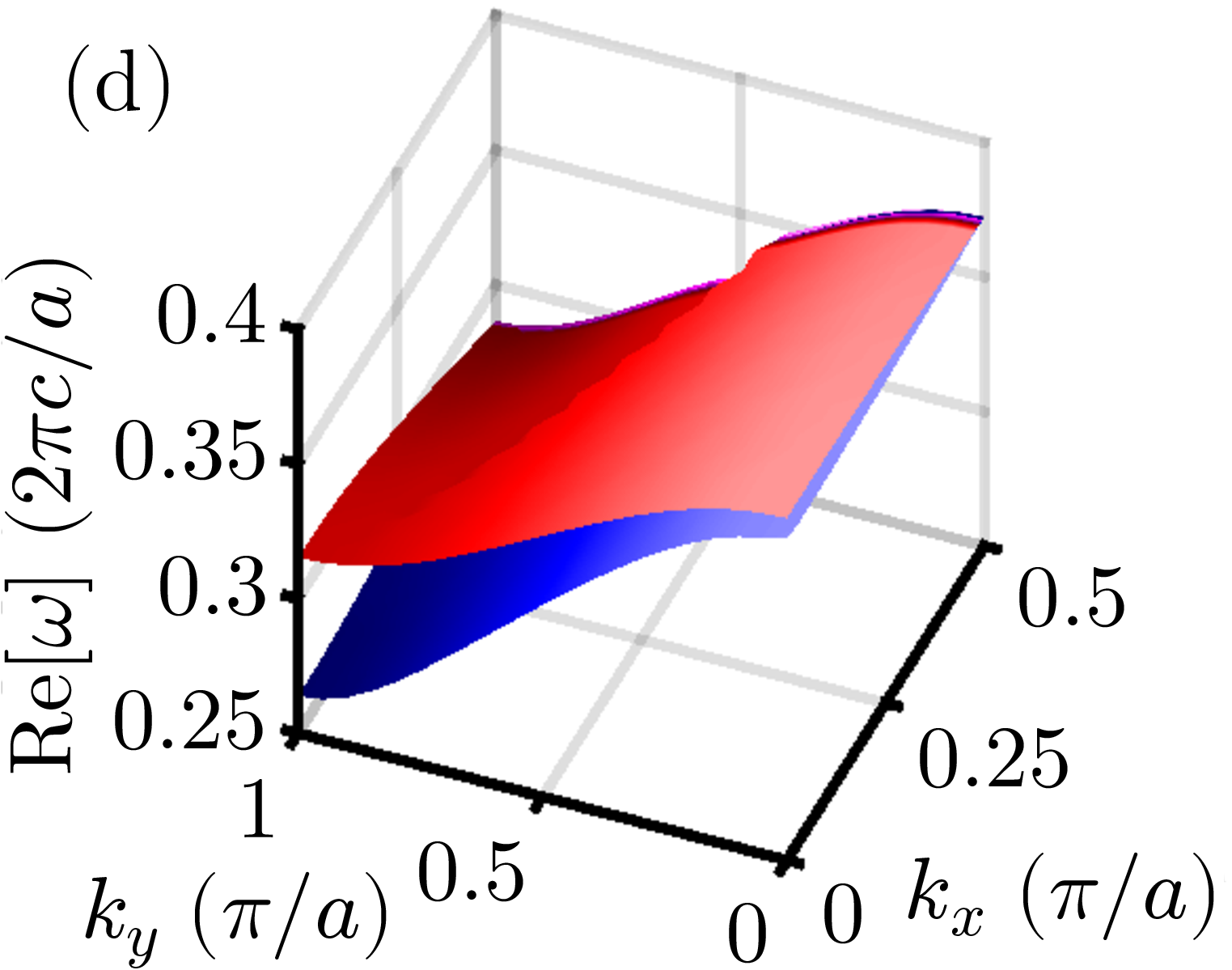}
    }
    \subfigure{
        \centering
        \includegraphics[width=0.22\textwidth]{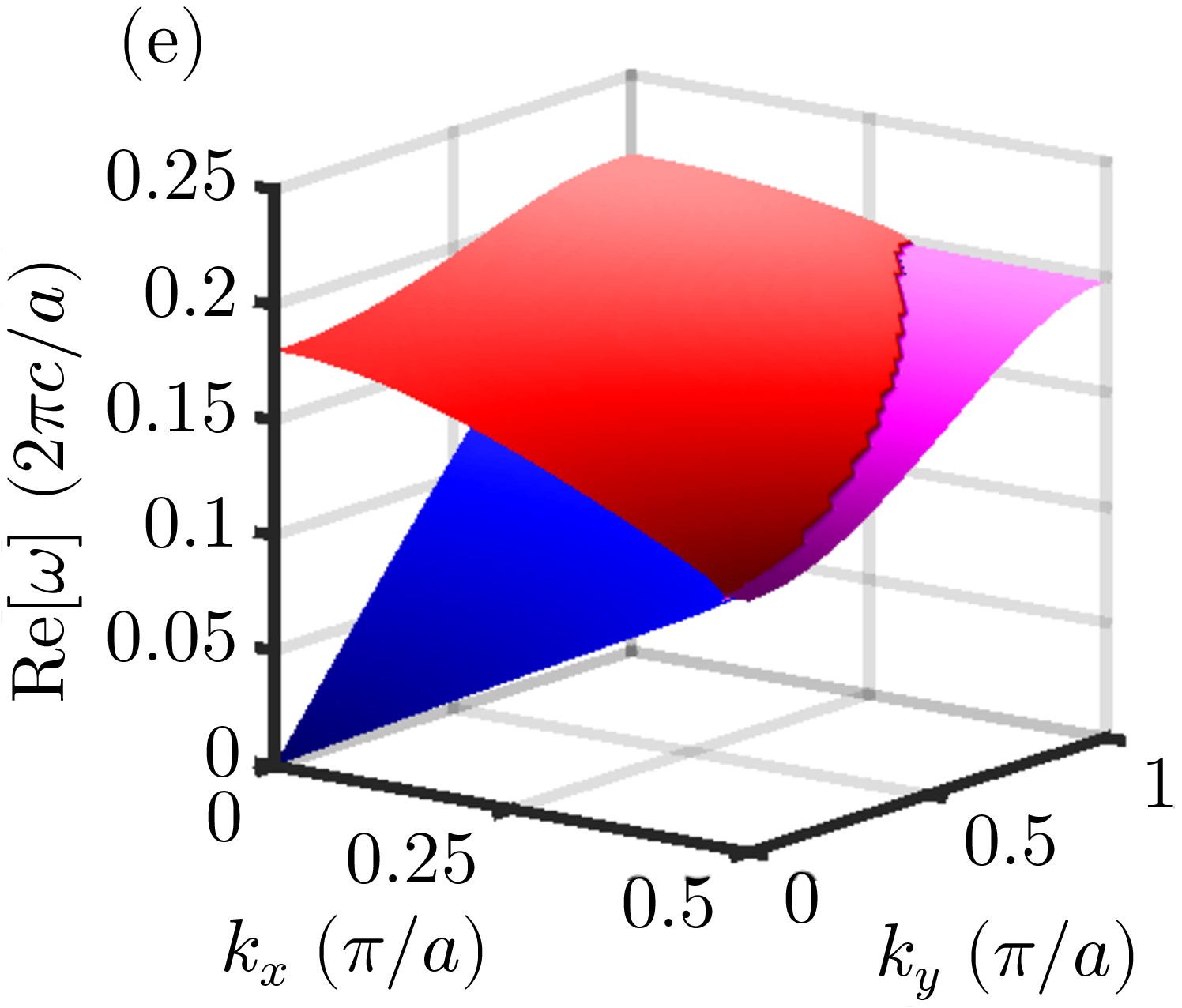}
        \includegraphics[width=0.22\textwidth]{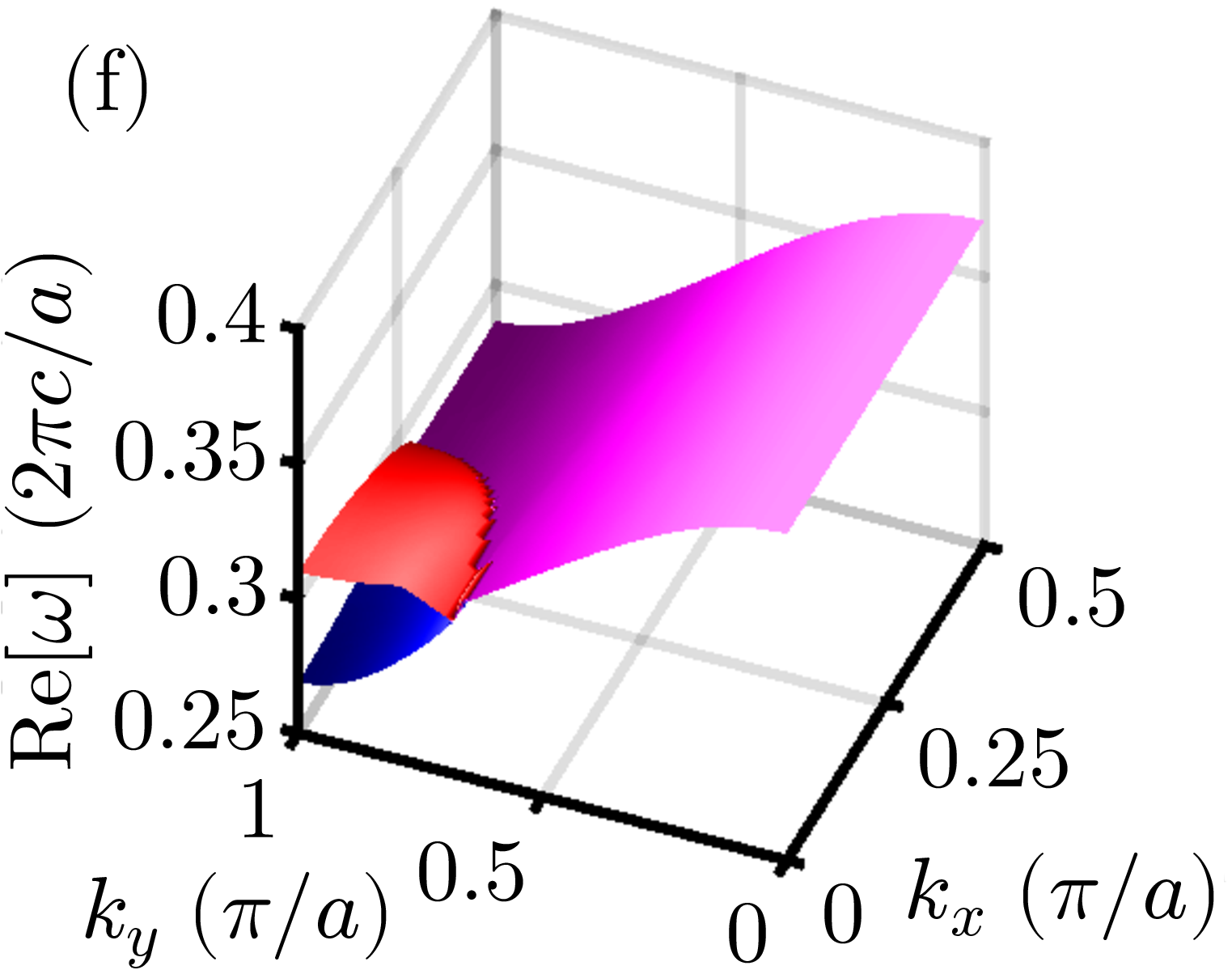}
    }
    \subfigure{
        \centering
        \includegraphics[width=0.22\textwidth]{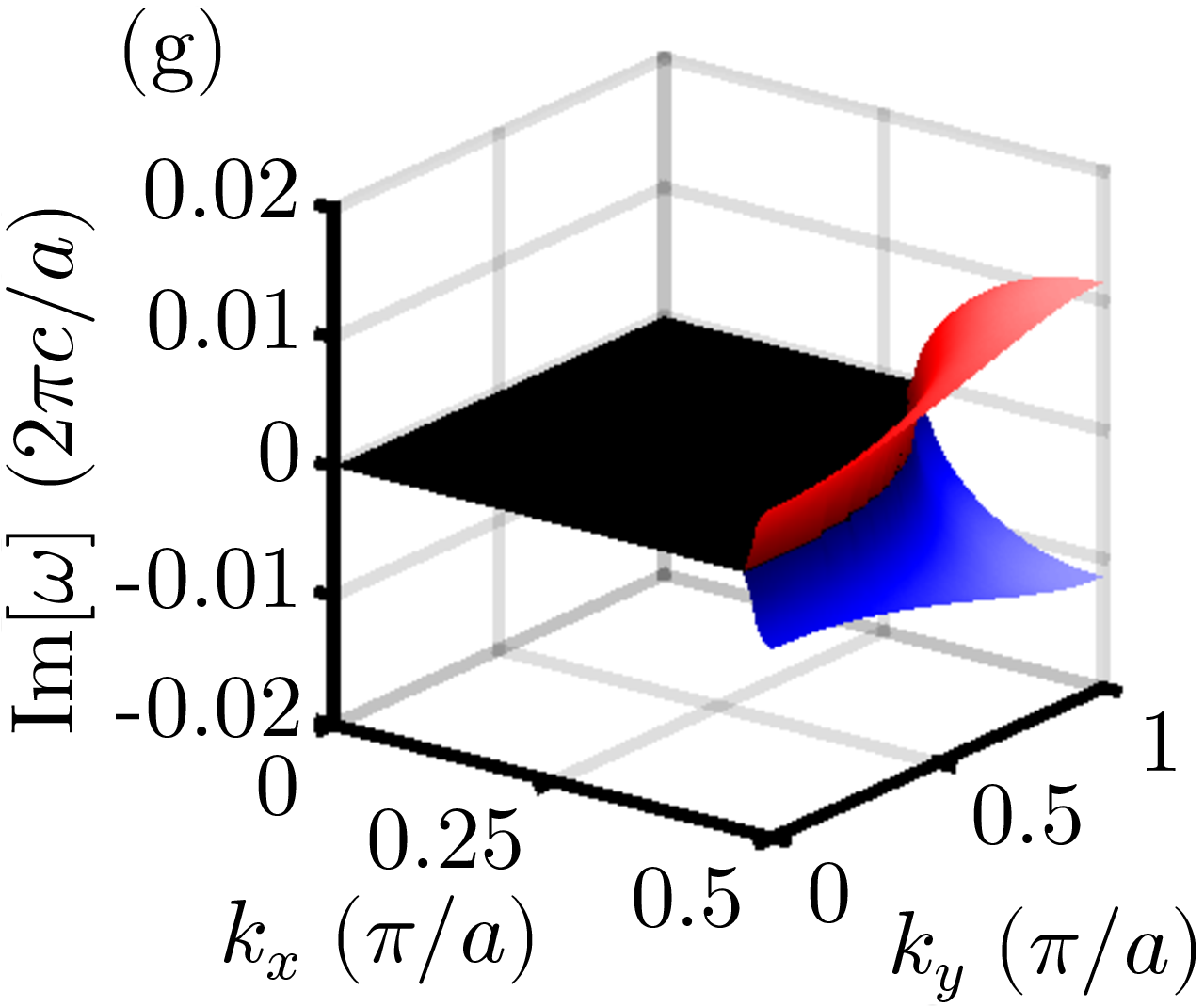}
        \includegraphics[width=0.22\textwidth]{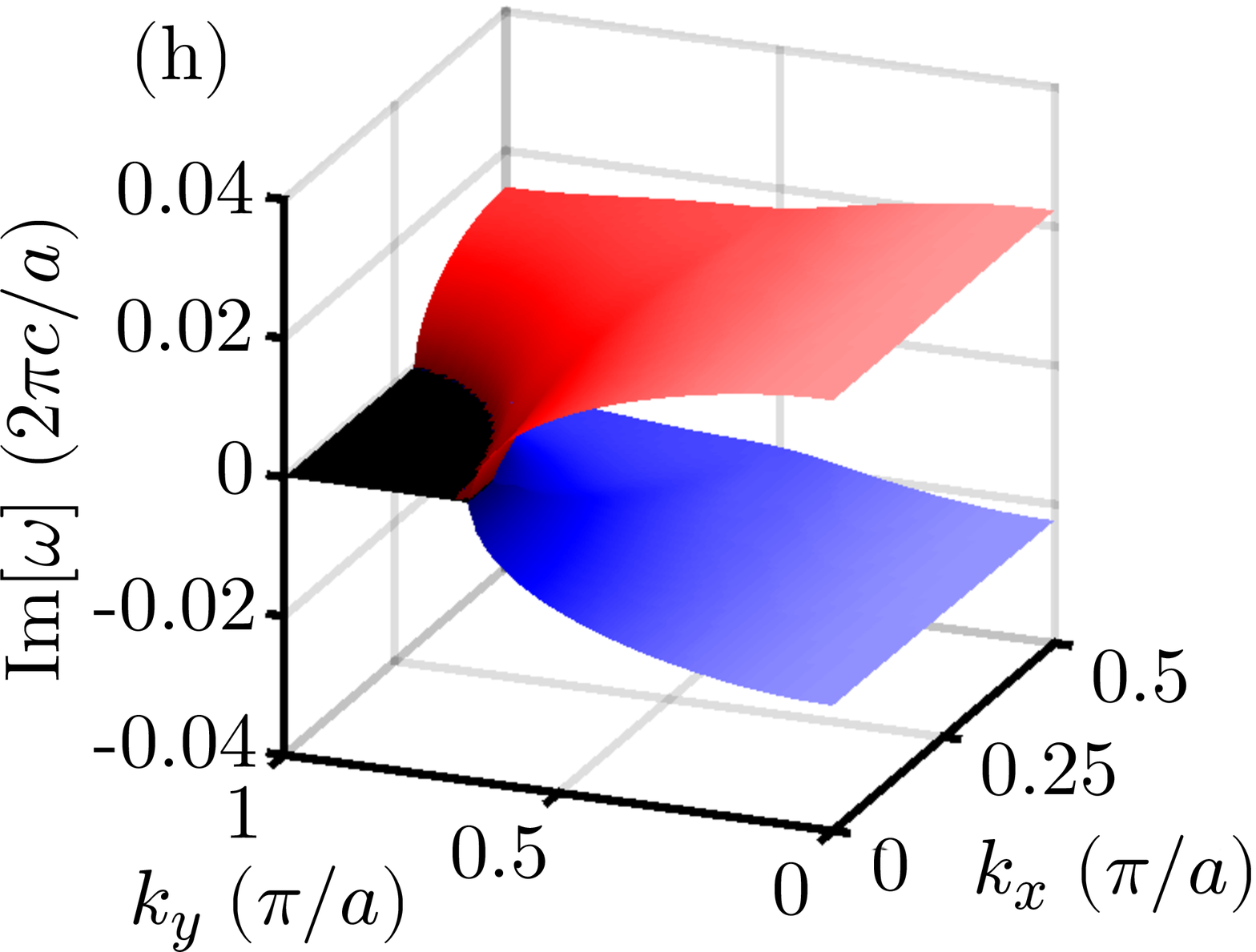}
    }
    \caption{(Color online) (a) Schematic of the 2D PhC comprised of square rods with side length $0.6a$ 
      of dielectric, $\varepsilon_{die} = 12$, embedded in air, $\varepsilon_{air} = 1$, with a square
      primitive cell side length of $a$. The primitive cell is indicated in gray, while the
      supercell contains two primitive cells and is marked with a dashed border. When $\tau \ne 0$,
      the red rods contain gain, while the cyan rods contain loss. (b) Schematic of the underlying 2D Hermitian PhC. 
      (c,e) Real part of the frequencies for the
      first (blue) and second (red) supercell TM bands when $\tau = 0$ and $\tau = 1.5$. Locations where
      the bands have merged are shown in magenta. (d,f) Real part of the frequencies for the fifth (blue) and sixth (red)
      supercell TM bands when $\tau = 0$ and $\tau = 1.5$. (g) Imaginary part of the frequencies for the
      first (blue) and second (red) supercell TM bands when $\tau = 1.5$. Black denotes no imaginary component. 
      (h) Imaginary part of the frequencies for the
      fifth (blue) and sixth (red) supercell TM bands when $\tau = 1.5$. 
      \label{fig:2x1}}

\vspace{-11pt}
\end{figure}

To illuminate this process, as an example we consider the two-dimensional PhC formed of dielectric
square rods with alternating gain or loss of equal magnitude embedded in air depicted in Fig.~\ref{fig:2x1}(a).
The primitive cell of this structure contains two square rods, one containing gain and a neighbor containing loss.
In the absence of gain and loss, the underlying Hermitian system, shown in Fig.~\ref{fig:2x1}(b), has a smaller primitive
cell containing a single dielectric rod. For semantic convenience, we will henceforth refer to this larger
primitive cell as the `supercell,' and reserve `primitive cell' for the underlying Hermitian system.

The band structure of the underlying Hermitian system, plotted with respect to the supercell, is shown
in Fig.~\ref{fig:2x1}(c) and (d) for the first and third sets of TM bands. We see that the band structure
is folded along the $k_x = \pi/2a$ line, where $a$ is the primitive cell lattice spacing, creating a degenerate contour. 
As gain and loss are added to the system, 
thresholdless $\mathcal{PT}$ transitions occur along these degenerate contours, while neighboring locations in wavevector space undergo ordinary $\mathcal{PT}$
transitions in a continuous manner, as can be seen in Figs.~\ref{fig:2x1}(e)-\ref{fig:2x1}(h). This causes
the folded bands to merge together outwards from the degenerate contour, forming the requisite
complex conjugate pairs of frequencies, while the boundary between the merged and independent regions
is a contour comprised entirely of exceptional points.
In the wake of this
folding process, the bands nearly flatten in the $x$-direction perpendicular to the degenerate contour.
However, unlike in 1D periodic structures \cite{sandhu_stopping_2007,ding_coalescence_2015}, this flattening does not correspond
to zero group velocity, as $\nabla_\mathbf{k} \omega$ can be non-zero in the $y$-direction of contiguous
gain or loss.





We now show that the behavior as observed above is generally present in $\mathcal{PT}$ symmetric
PhCs that have an enlarged primitive cell (i.e.~a `supercell' as defined above) as compared to the
underlying Hermitian system (i.e.~a `primitive cell' as defined above).
The band structure of a $\mathcal{PT}$ symmetric PhC is defined by
\begin{equation}
\left[ \nabla \times \nabla \times - \left(\varepsilon(\mathbf{x}) + i \tau g(\mathbf{x}) \right) \frac{\omega_n^2(\mathbf{k})}{c^2} \right] \mathbf{E}_{n\mathbf{k}}(\mathbf{x}) = 0, \label{eq:phcWave}
\end{equation}
in which $\mathbf{E}_{n\mathbf{k}}(\mathbf{x})$ is the mode profile of the
$n$th band with wavevector $\mathbf{k}$ and frequency $\omega_n(\mathbf{k})$,
$\varepsilon(\mathbf{x})$ is the Hermitian dielectric function
of the PhC, and $\tau$ and $g(\mathbf{x})$ are the strength and
distribution of the gain and loss in the PhC respectively. 
We assume that the primitive cell has a set of lattice vectors $\{ \mathbf{a} \}$ such
that $\varepsilon(\mathbf{x}+\mathbf{a}_i) = \varepsilon(\mathbf{x})$,
while the supercell has lattice vectors $\{ \mathbf{A} \}$,
which are usually integer multiples of the primitive cell lattice vectors,
$\mathbf{A}_i = \sum_j n_{ij} \mathbf{a}_j$, such that $g(\mathbf{x} + \mathbf{A}_i) = g(\mathbf{x})$.
We also define the supercell's reciprocal lattice vectors $\mathbf{B}_i$,
such that $\mathbf{B}_i \cdot \mathbf{A}_j = 2\pi \delta_{ij}$.
Given the periodicity of the supercell, the mode profiles obey the supercell translation symmetry, 
\begin{equation}
\mathbf{E}_{n\mathbf{k}}(\mathbf{x}+ \mathbf{A}_j)= e^{i\mathbf{k}\cdot \mathbf{A}_j}\mathbf{E}_{n\mathbf{k}}(\mathbf{x}).
\end{equation}
Finally, the gain and loss are applied such that $g(\mathbf{x}) = -g(-\mathbf{x})$, and
we adopt the convention that $\tau \ge 0$.


Our objective is to explore the behavior of the band structure in the vicinity of a particular
set of $\mathbf{k}$-points such as the degenerate contour. Thus, for a region of $\mathbf{k}$-space
in the neighborhood of a wavevector $\mathbf{k}_0$, we expand the supercell wavefunctions at
$\mathbf{k}$ in terms of those at $\mathbf{k}_0$ as
\begin{equation}
\mathbf{E}_{n\mathbf{k}}(\mathbf{x}) = \sum_m C_{nm}(\mathbf{k}) e^{i(\mathbf{k}-\mathbf{k}_0)\cdot \mathbf{x}} \mathbf{E}_{m\mathbf{k}_0}^{(0)}(\mathbf{x}), \label{eq:eExp}
\end{equation}
where $\mathbf{E}_{m\mathbf{k}}^{(0)}(\mathbf{x})$ satisfies Eq.~(\ref{eq:phcWave}) with $\tau = 0$,
and $C_{nm}$ are the complex expansion coefficients \cite{johnson_k.p_1993,johnson_theory_1994}. 
In doing so, we avoid the difficulties associated
with using the Bloch modes of a non-Hermitian structure \cite{makris_beam_2008,longhi_bloch_2009,makris_pt-symmetric_2010},
and can normalize the wavefunctions in the usual manner,
\begin{equation}
\int_{\textrm{SC}} \varepsilon(\mathbf{x}) \left(\mathbf{E}_{n\mathbf{k}}^{(0)}(\mathbf{x}) \right)^* \cdot \mathbf{E}_{m\mathbf{k}'}^{(0)}(\mathbf{x}) d\mathbf{x} = \delta_{nm} \delta(\mathbf{k}-\mathbf{k}'),
\end{equation}
where the integral is evaluated over the supercell.
Furthermore, since the supercell with $\tau = 0$ is an exact $N$-fold copy of the primitive cell,
there are exactly $N$ reciprocal lattice vectors that are integer multiples of the members of $\{ \mathbf{B} \}$ which
generate the primitive Brillouin zone from the supercell Brillouin zone, and are denoted as $\mathbf{L}_1,...,\mathbf{L}_N$, such that $\mathbf{L}_j = \sum_i m_i \mathbf{B}_i$. %
Thus, as the translational symmetry of the underlying Hermitian system is described by the primitive cell,
the states $\mathbf{E}_{m\mathbf{k}}^{(0)}(\mathbf{x})$ satisfy a `hidden' translational symmetry,
and can be chosen such that
\begin{equation}
\mathbf{E}_{m\mathbf{k}}^{(0)}(\mathbf{x}+\mathbf{a}_j) = e^{i(\mathbf{k}+\mathbf{L}_i)\cdot \mathbf{a}_j} \mathbf{E}_{m\mathbf{k}}^{(0)}(\mathbf{x}). \label{eq:hidden}
\end{equation}
Here, each supercell band $m$ only satisfies this relationship for a single element $\mathbf{L}_j$ \cite{allen_recovering_2013}.
Furthermore, each of the supercell bands which correspond to the same unfolded band from the primitive Brillouin
zone satisfies Eq.~(\ref{eq:hidden}) for a different $\mathbf{L}$. Thus, one could re-index
the supercell bands as $m = (\nu,j)$, where $\nu$ is the index of the corresponding band of the primitive Brillouin zone,
and the wavefunction satisfies Eq.~(\ref{eq:hidden}) for $\mathbf{L}_j$.

Upon substituting
Eq.~(\ref{eq:eExp}) into Eq.~(\ref{eq:phcWave}), multiplying through by $(\mathbf{E}_{l\mathbf{k}_0}^{(0)}(\mathbf{x}))^*$, and
integrating over the supercell, we find the matrix equation
\begin{multline}
\sum_m \Big[ \left(\omega_n^2(\mathbf{k}) - (\omega_m^{(0)}(\mathbf{k}_0))^2 \right) \frac{\delta_{lm}}{c^2} + i \tau \frac{\omega_n^2(\mathbf{k})}{c^2} G_{lm} \\
   + \mathbf{s}\cdot \mathbf{P}_{lm} - s^2 Q_{lm} \Big] C_{nm}(\mathbf{k}) = 0, \label{eq:kdotp1}
\end{multline}
where $\mathbf{s} = \mathbf{k} - \mathbf{k}_0$ and $\omega_m^{(0)}(\mathbf{k}_0)$ is the frequency of the
$m$th band of the supercell system when $\tau=0$. For ease of the following analysis we have specialized to 2D TM bands, but
a full vectorial treatment is straightforward \cite{supp_matt}.
The matrix element $G_{lm}$ contains the effects
of modal coupling through the gain and loss, 
\begin{equation}
G_{lm} = \int_{\textrm{SC}} g(\mathbf{x}) \left(E_{l\mathbf{k}_0}^{(0)}(\mathbf{x}) \right)^* E_{m\mathbf{k}_0}^{(0)}(\mathbf{x}) d\mathbf{x},
\end{equation}
while the elements $\mathbf{P}_{lm}$ and $Q_{lm}$
represent the frequency shifts due to displacements in $\mathbf{k}$-space for the Hermitian
system,
\begin{align}
\mathbf{P}_{lm} =& 2i\int_{\textrm{SC}} \left(E_{l\mathbf{k}_0}^{(0)}(\mathbf{x}) \right)^* \nabla E_{m\mathbf{k}_0}^{(0)}(\mathbf{x}) d\mathbf{x}, \\
Q_{lm} =& \int_{\textrm{SC}} \left(E_{l\mathbf{k}_0}^{(0)}(\mathbf{x}) \right)^* E_{m\mathbf{k}_0}^{(0)}(\mathbf{x})  d\mathbf{x}.
\end{align}
Note, the group velocity for each band is given by the corresponding diagonal element of
$\mathbf{P}$ as $\nabla_{\mathbf{k}} \omega_m(\mathbf{k}_0) = -c^2 \mathbf{P}_{mm}/2 \omega_m(\mathbf{k}_0)$,
which is true even for locations with degenerate frequencies due to the requirement that $\mathbf{E}_{m\mathbf{k}_0}^{(0)}(\mathbf{x})$ satisfy Eq.~(\ref{eq:hidden}).
Furthermore, the hidden translational symmetry of the wave functions of the Hermitian supercell system yields two important
restrictions upon the coupling matrix elements. First, it can be shown that by breaking up the
integrals over the supercell into the individual primitive cell constituents, $\mathbf{P}_{lm}$ and $Q_{lm}$
are only non-zero if $\mathbf{E}_{l\mathbf{k}_0}^{(0)}(\mathbf{x})$ and $\mathbf{E}_{m\mathbf{k}_0}^{(0)}(\mathbf{x})$
obey Eq.~(\ref{eq:hidden}) for the same $\mathbf{L}_j$ \cite{supp_matt}. Second, the odd parity
symmetry of $g(\mathbf{x})$ results in $G_{mm} = 0$. As the wave functions for any point in $\mathbf{k}$-space form a complete set,
Eq.~(\ref{eq:kdotp1}) is an exact restatement of Eq.~(\ref{eq:phcWave}) 
(although extra considerations are necessary for vectorial fields \cite{sipe_vector_2000}).

The application of gain and loss to the Hermitian system couples pairs of bands
in the supercell system which originate from the same unfolded band of the primitive
Brillouin zone. Thus, we will assume that we can decouple any such pair of bands from the rest of the system,
and rewrite Eq.~(\ref{eq:kdotp1}) for the reduced two-band system as
\begin{equation}
\frac{\omega^2(\mathbf{k})}{c^2} \left[ 
\begin{array}{cc}
1  & i\tau G_{12} \\
i\tau G_{21} & 1 
\end{array} \right] \mathbf{C} = \Omega(\mathbf{s},\mathbf{k}_0) \mathbf{C}, \label{eq:2lv}
\end{equation}
where $\Omega_{ij}(\mathbf{s},\mathbf{k}_0) = [(\omega_i^{(0)}(\mathbf{k}_0)/c)^2 - \mathbf{s}\cdot \mathbf{P}_{ii} + s^2 Q_{ii}]\delta_{ij}$. 
By setting $\mathbf{s}=0$, the non-Hermitian PhC satisfies Eq.~(\ref{eq:2lv}) over all of $\mathbf{k}$-space, 
and Eq.~(\ref{eq:2lv}) correctly reduces to Eq.~(3) of Ge and Stone for systems with isolated modes \cite{ge_parity-time_2014},
except that Eq.~(\ref{eq:2lv}) has been derived for systems with continuous bands.

However, in contrast to previous works \cite{ge_parity-time_2014,ding_coalescence_2015}, we can now
choose $\mathbf{k} \ne \mathbf{k}_0$ to understand the the band merging process.
To this end, we select $\mathbf{k}_0$ to be a degenerate point of the supercell Hermitian system with frequency $\omega^{(0)}(\mathbf{k}_0)$, 
and solve for the frequencies of the non-Hermitian system as
\begin{equation}
\frac{\omega^2}{c^2} = \frac{2\Omega_{11}\Omega_{22}}{\Omega_{11}+\Omega_{22} \pm \sqrt{(\Omega_{11}-\Omega_{22})^2 - 4\Omega_{11}\Omega_{22}\tau^2|G_{12}|^2} }.
\end{equation}
As the two supercell bands originate from the same primitive band, the association of $\mathbf{P}_{ii}$ with the group velocity yields two related
conclusions. First, along the degenerate contour, $ \partial \omega_1(\mathbf{k}_0) /\partial k_\parallel = \partial \omega_2(\mathbf{k}_0) /\partial k_\parallel = -c^2 P_\parallel / 2 \omega^{(0)}$, while
perpendicular to the degenerate contour, $ \partial \omega_2(\mathbf{k}_0)/\partial k_\perp = - \partial \omega_1(\mathbf{k}_0) /\partial k_\perp = -c^2 P_\perp / 2 \omega^{(0)}$, as the unfolded
band of the primitive cell is smooth. Thus, the threshold for $\mathcal{PT}$ symmetry breaking to second order in $\mathbf{s}$ is
\begin{equation}
\tau_{\textrm{TH}} \approx \left| \frac{s_\perp P_{\perp}\left(1 + \frac{c^2 s_\parallel P_{\parallel}}{(\omega^{(0)}(\mathbf{k}_0))^2}\right)+ \frac{s^2}{2}(Q_{22}-Q_{11})}
{|G_{12}|\left(\frac{\omega^{(0)}(\mathbf{k}_0)}{c}\right)^2} \right|. \label{eq:tauPred}
\end{equation}
To first order, $\tau_{\textrm{TH}}$ is seen to be strictly dependent upon the perpendicular displacement
in wavevector space from the degenerate contour, in agreement with the band structures seen in Figs.~\ref{fig:2x1}(e) and \ref{fig:2x1}(f).
Furthermore, the second order corrections yield an increase
in the threshold calculated about a particular $\mathbf{k}_0$ if $\mathbf{s}$ also contains a component
parallel to the degenerate contour. Thus,
the correct (minimum) $\mathcal{PT}$ threshold for any point $\mathbf{k}$ is calculated from the closest location on the degenerate contour,
and is seen to be strictly dependent upon $s_\perp$, demonstrating
that the coupled bands of the non-Hermitian system merge together directly outwards from the degenerate
contour continuously.

The flattening of the bands as they merge can be understood by solving for the frequency at the exceptional
point (using $\mathbf{s} = s_\perp$),
\begin{equation}
\omega_{\textrm{TH}}(\mathbf{k}) \approx \omega^{(0)}(\mathbf{k}_0) \left(1 + \frac{c^2 s_\perp^2 (Q_{11}+Q_{22})}{4(\omega^{(0)}(\mathbf{k}_0))^2}\right), \label{eq:flat}
\end{equation}
which is seen to be given by the frequency of the associated degenerate frequency with the leading correction
being second order in $s_\perp c / \omega^{(0)}$. 
As $\tau$ is increased beyond the threshold value 
for a particular location in wavevector space, the dominant change in the frequencies of the two bands is to acquire
imaginary components, with only minor shifts in the real components, which leaves the bands nearly flat after they merge.


\begin{figure}[!t]
    \centering
    \subfigure{
    \centering
        \includegraphics[width=0.28\textwidth]{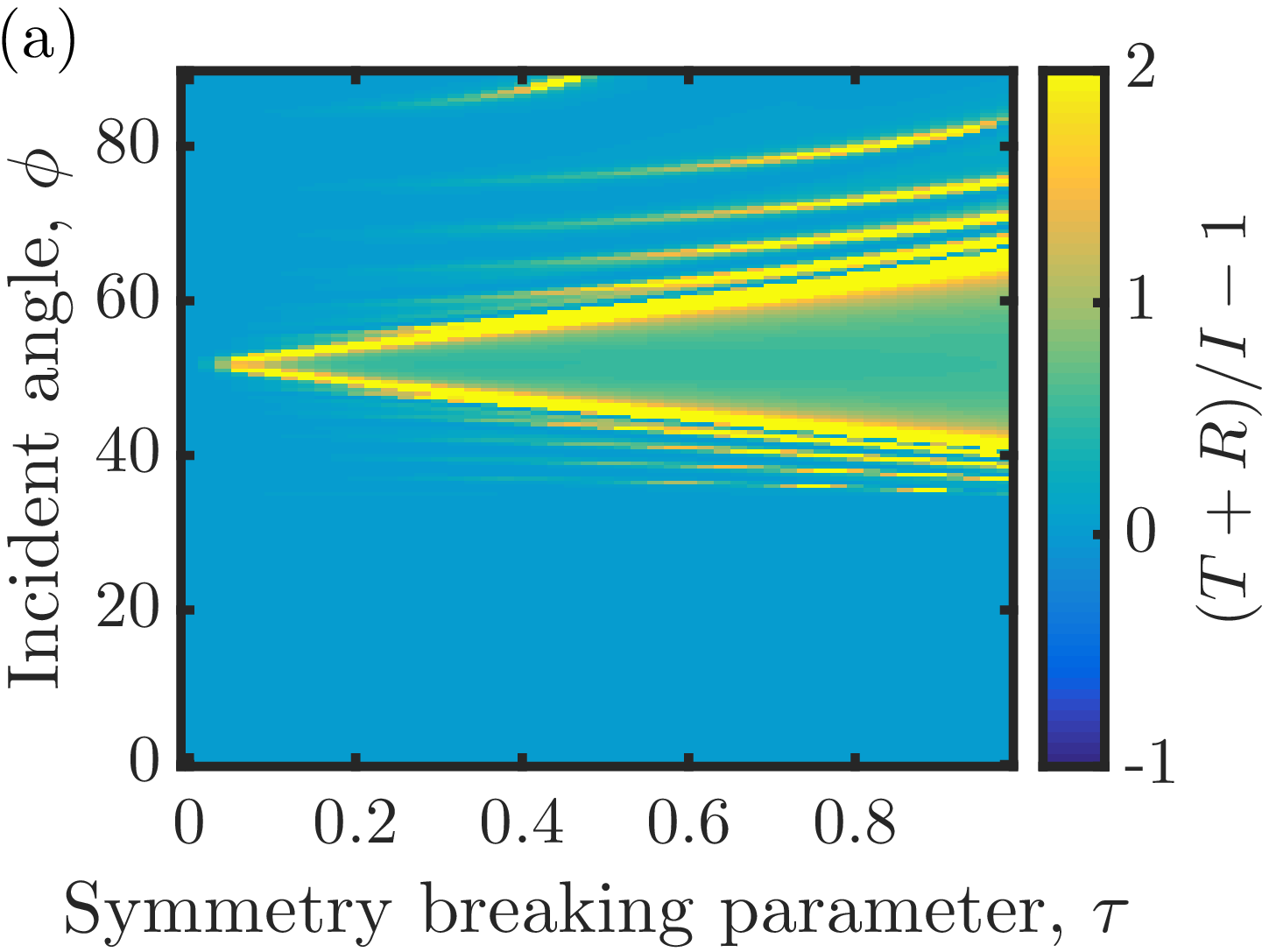}
        \includegraphics[width=0.08\textwidth]{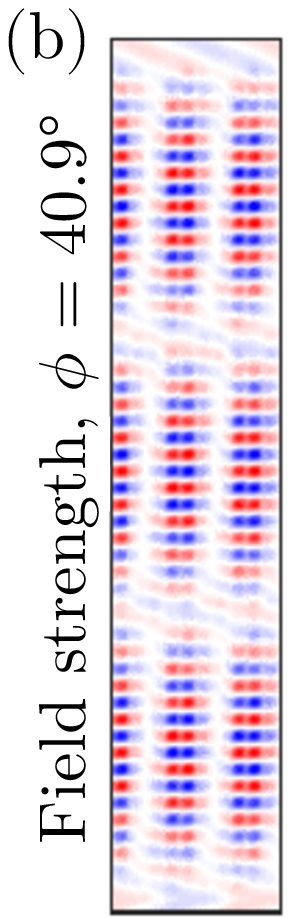}
        \includegraphics[width=0.08\textwidth]{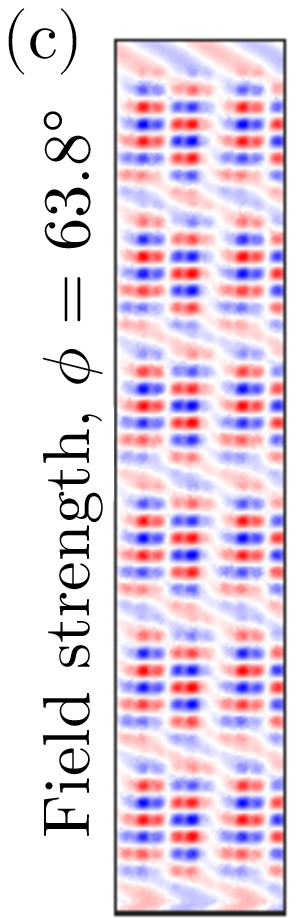}
    }
    \caption{(Color online) (a) Non-unitary behavior as a function of
      incident angle, $\phi$, and $\mathcal{PT}$ symmetry breaking parameter, $\tau$, for an $s$-polarized plane wave source with $\omega = 0.185(2\pi c / a)$ incident upon a PhC
      slab infinite in the $x$-direction and with $50$ layers in the $y$-direction for the same
      system shown in Fig.~\ref{fig:2x1}(a). $\phi = 0$ corresponds to normal incidence in the $y$-direction. The PhC slab is surrounded by a passive dielectric with $\varepsilon = 3$.
      Values of $0$ correspond to unitary behavior, while
      $[-1,0)$ signifies absorption, and $(0,\infty]$ signifies amplification.
      The reflection and transmission coefficients were calculated using the Fourier Modal Method as implemented in S$^4$ \cite{liu_S4}.
      (b-c) Plot of the real part of the electric field for the same structure with $\tau = 0.65$, and $\phi = 40.9^\circ$ (c),
      or $\phi = 63.8^\circ$ (d). Field plots were generated using the freely available MaxwellFDFD software package \cite{shin_maxwellfdfd}. 
      \label{fig:superprism}}

\vspace{-11pt}
\end{figure}

The thresholdless $\mathcal{PT}$ transition of supercell $\mathcal{PT}$ symmetric PhCs can also be observed in the non-unitary behavior of
related finite systems. Figure \ref{fig:superprism}(a) shows the amplification and absorption as a function
of incident angle and $\tau$ for a single frequency incident upon a PhC slab similar to Fig.~\ref{fig:2x1}(a), which is infinite in the
$x$-direction, but finite in the $y$-direction. The frequency chosen lies within the range of frequencies comprising the degenerate
contour of the first pair of bands, Fig.~\ref{fig:2x1}(e), and thus should exhibit a thresholdless $\mathcal{PT}$ transition
at a particular incidence angle.
As $\tau$ is increased, the area of the $\mathcal{PT}$-broken
region in $\mathbf{k}$-space is increased, resulting in a wider range of incident angles which yield amplification.

In Hermitian PhCs, the superprism effect refers to sharp features in the isofrequency contours
of a band structure, where a small change in the incident angle of light yields an enormous
change in the refraction angle of the light inside the PhC \cite{joannopoulos,kosaka_superprism_1998,kosaka_superprism_1999}. However, the
exceptional contour of a $\mathcal{PT}$ PhC separates a region of non-Hermitian behavior
from that of ordinary propagation. This yields a `$\mathcal{PT}$-superprism' effect, in which
a small change in the incident angle of the signal results in either unitary or non-unitary
behavior. For example, when $\tau = 0.65$, the system is unitary at $\phi = 39^\circ$,
and yet exhibits a tenfold increase in the net gain with the small change of the incidence angle to $\phi = 44^\circ$.
This effect could have applications as an optical switch.

\begin{figure}[!t]
    \centering
    \subfigure{
    \centering
        \includegraphics[width=0.23\textwidth]{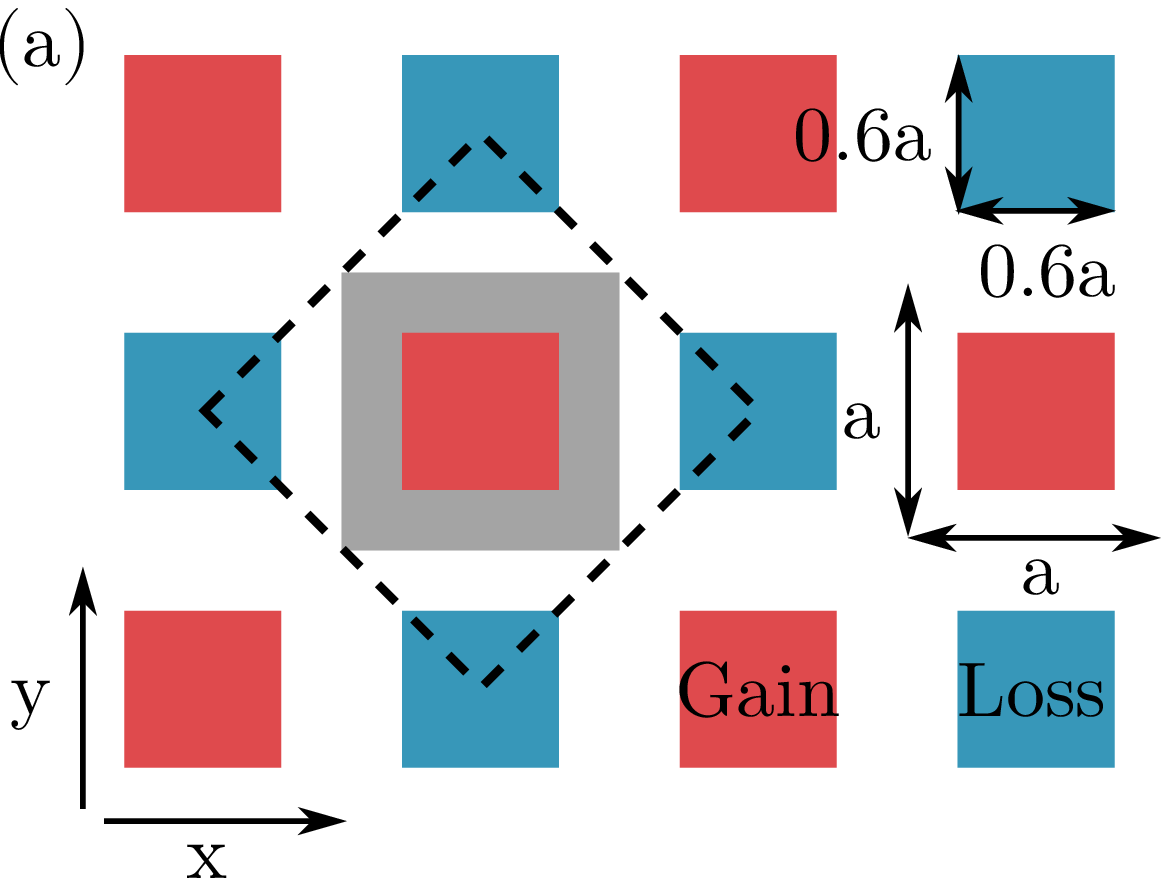}
        \includegraphics[width=0.22\textwidth]{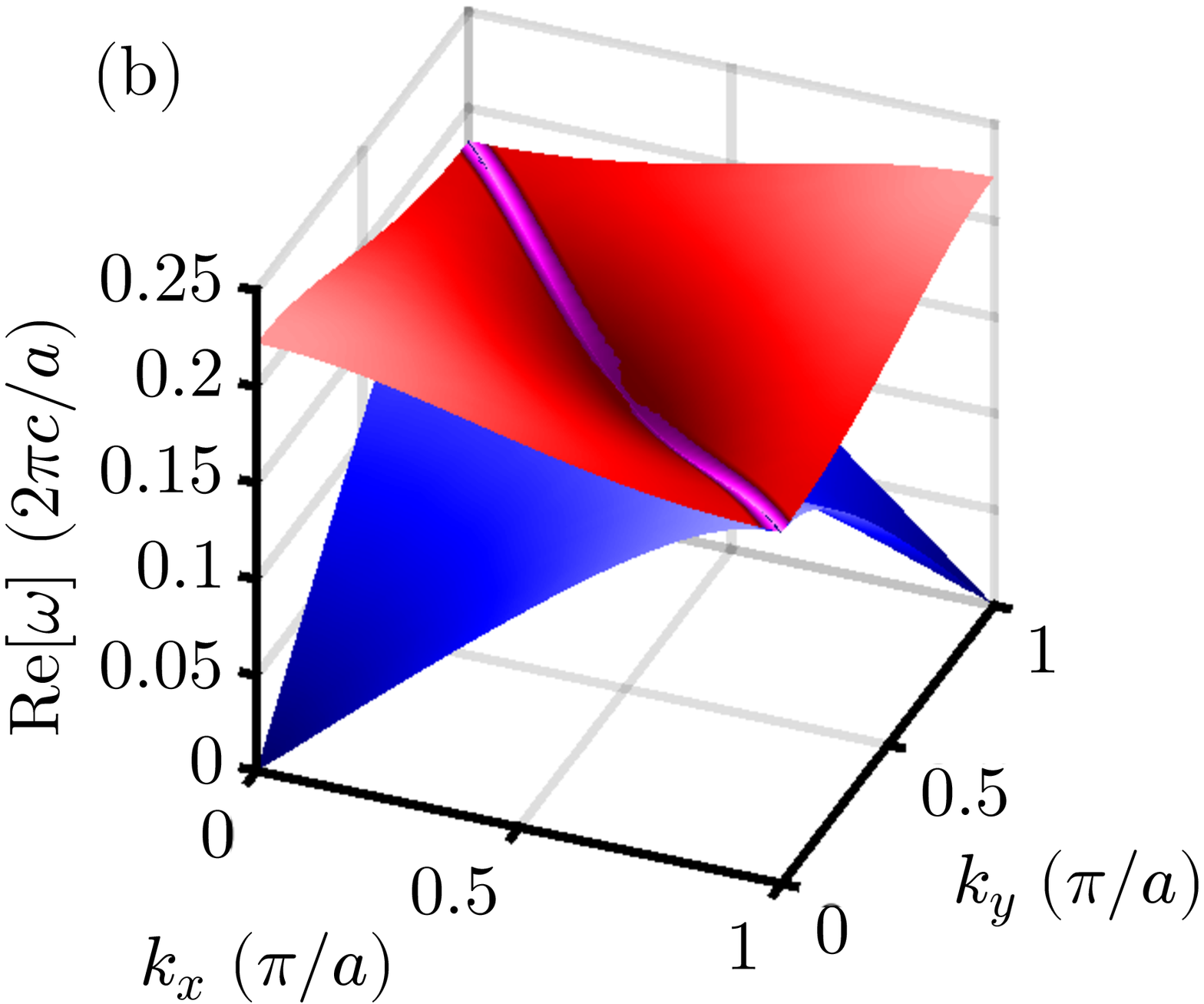}
    }
    \subfigure{
    \centering
        \includegraphics[width=0.22\textwidth]{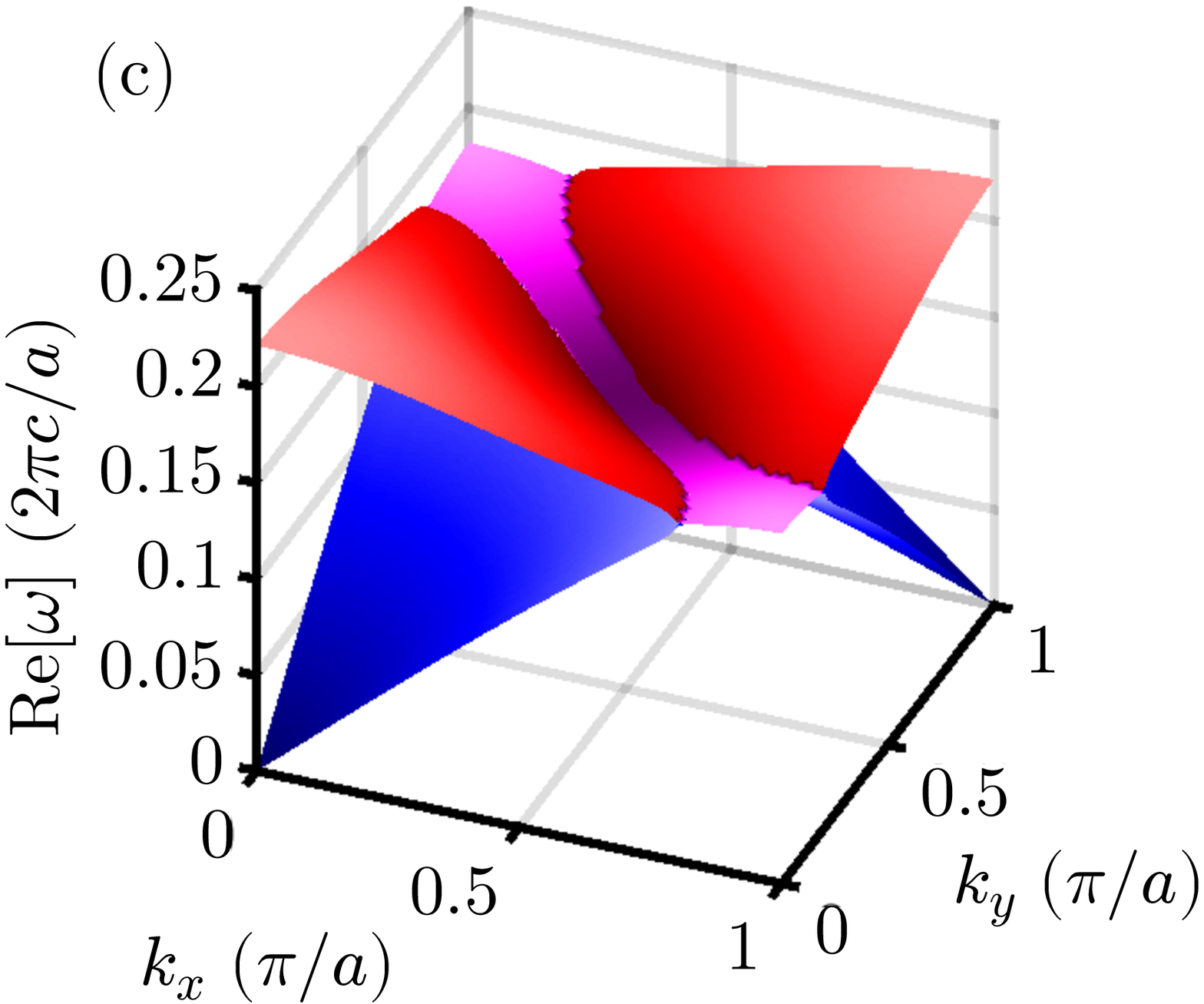}
        \includegraphics[width=0.22\textwidth]{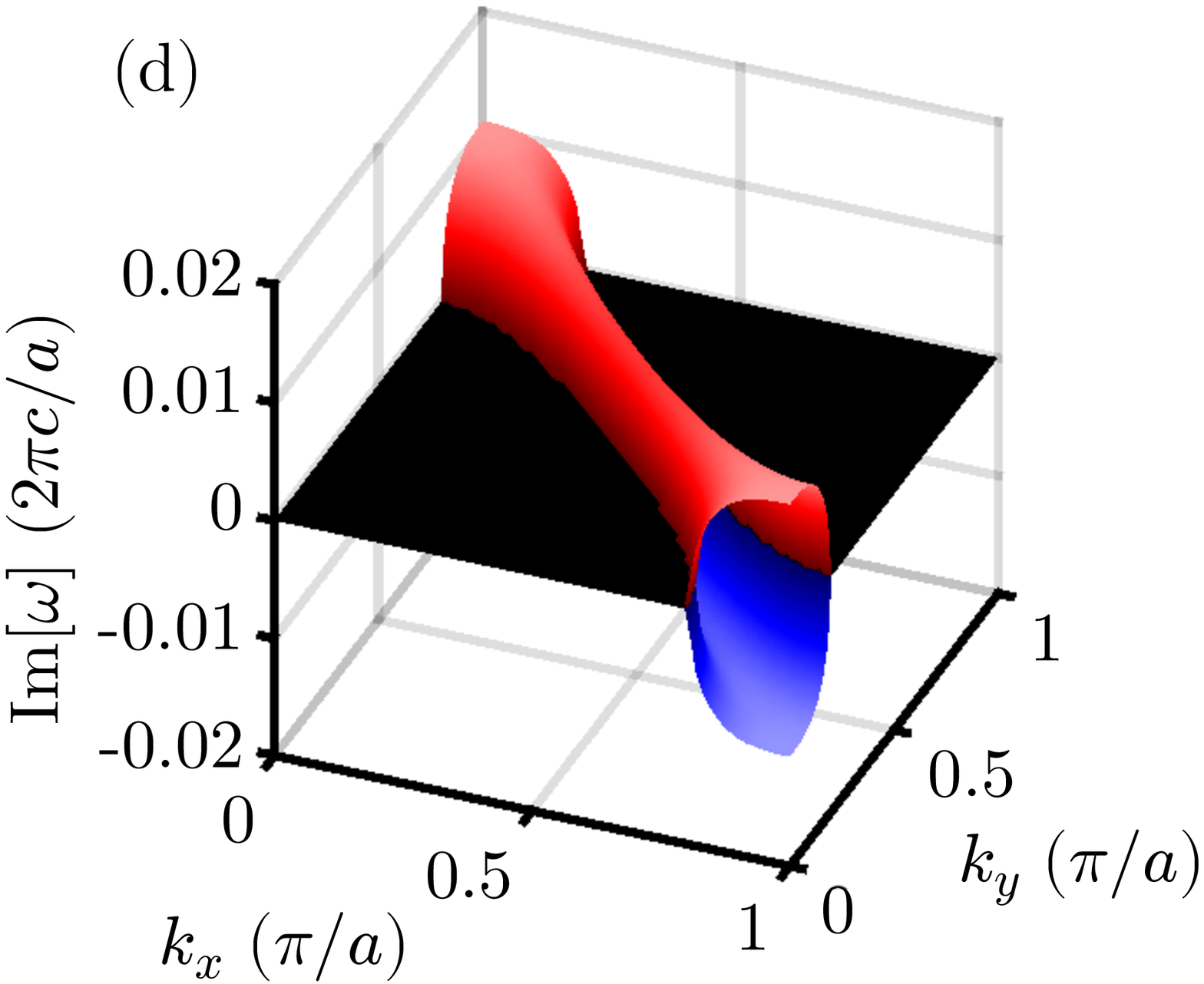}
    }
    \caption{(Color online) (a) Schematic of the 2D PhC comprised of square rods with side length $0.6a$ 
      of dielectric, $\varepsilon_{die} = 12$, embedded in air, $\varepsilon_{air} = 1$, with a square
      primitive cell side length of $a$. The primitive cell is indicated in gray, while the
      supercell contains two primitive cells and is marked with a dashed border. When $\tau \ne 0$,
      the red rods contain gain, while the cyan rods contain loss. (b,c) Real part of the frequencies for the
      first (blue) and second (red) supercell TM bands when $\tau = 0$ and $\tau = 1.5$. Locations where
      the bands have merged are shown in magenta. (d) Imaginary part of the frequencies for the
      first (blue) and second (red) supercell TM bands when $\tau = 1.5$. Black denotes no imaginary component. 
      \label{fig:2x2}}

\vspace{-11pt}
\end{figure}

Likewise, flat features in isofrequency contours act as supercollimators,
counteracting diffraction for incident beams with a finite width whose Fourier components lie
within the flat contour \cite{joannopoulos,luo_all-angle_2002,yu_bends_2003}. As is seen in the band-merging process in Figs.~\ref{fig:2x1}(c)-\ref{fig:2x1}(h),
by changing $\tau$, flat contours can be designed with a desired width, potentially spanning the entire Brillouin zone, or removed entirely,
allowing for tunable supercollimation or all-angle supercollimation for frequencies with
completely merged bands. 
The effect of supercollimation can also be
seen in the field profiles of the finite PhC system, where within the $\mathcal{PT}$-broken
region the wavefunction propagates entirely in the $y$-direction, 
as can be seen in the field profiles at the edge of this region in
Fig.~\ref{fig:superprism}(b) and \ref{fig:superprism}(c).
Finally, $\mathcal{PT}$ PhCs can exhibit unidirectional behavior \cite{supp_matt}.


By changing the distribution of gain and loss in the system while maintaining $\mathcal{PT}$ symmetry,
we can change the location of the degenerate contour in $\mathbf{k}$-space. An example of this is
shown in Fig.~\ref{fig:2x2}(a), where same underlying PhC from Fig.~\ref{fig:2x1}(b) is
considered with a different application of gain and loss. The degenerate contour of the supercell
Hermitian system now lies along the $X$-$Y$ contour of the primitive Brillouin zone, Fig.~\ref{fig:2x2}(b),
and as $\tau$ is increased, the $\mathcal{PT}$-broken region is seen to expand away from this contour, Figs.~\ref{fig:2x2}(c) and \ref{fig:2x2}(d).
This enables a new form of
band structure engineering, both by being able to choose an arbitrary contour to be the degenerate
contour, and through the possibility of electrically altering the distribution of gain and loss,
allowing for qualitative changes in the optical properties of the PhC without refabricating the PhC.

In conclusion, we have demonstrated that the degeneracies naturally generated in supercell $\mathcal{PT}$ symmetric PhCs 
can yield new control over band structure design. Furthermore, $\mathcal{PT}$ symmetric PhCs can exhibit qualitatively
new behaviors, such as the $\mathcal{PT}$-superprism effect, and all-angle supercollimation.


\begin{acknowledgments}
We would like to thank Jessica Piper, Li Ge, and Steven G.\ Johnson for helpful discussions.
This work was supported by the AFOSR MURI program (Grant No.\ FA9550-12-1-0471). 
\end{acknowledgments}


%

\end{document}